# Wavetable Synthesis Using CVAE for Timbre Control Based on Semantic Label

Tsugumasa Yutani*, Yuya Yamamoto*, Shuyo Nakatani†, Hiroko Terasawa*
* University of Tsukuba, Tsukuba, Japan
E-mail: tsugumasa320@gmail.com, s2130507@u.tsukuba.ac.jp, terasawa@slis.tsukuba.ac.jp
† Cybozu Labs, Tokyo, Japan
E-mail: nakatani@labs.cybozu.co.jp

*Abstract*—Synthesizers are essential in modern music production. However, their complex timbre parameters, often filled with technical terms, require expertise. This research introduces a method of timbre control in wavetable synthesis that is intuitive and sensible and utilizes semantic labels. Using a conditional variational autoencoder (CVAE), users can select a wavetable and define the timbre with labels such as $bright$, $warm$, and $rich$. The CVAE model, featuring convolutional and upsampling layers, effectively captures the wavetable nuances, ensuring real-time performance owing to their processing in the time domain. Experiments demonstrate that this approach allows for real-time, effective control of the timbre of the wavetable using semantic inputs and aims for intuitive timbre control through data-based semantic control. Audio examples are available on the online demo page[1].

## I. Introduction

Synthesizers, electronic musical instruments that create sounds using analog or digital signals, play a significant role in modern music production and performance. Although various methods of synthesizing sounds exist, effectively controlling the synthesis algorithm and obtaining the desired timbre requires specialized knowledge and experience. This complexity arises because numerous synthesis parameters are intricate, low-level, and unrelated to the perceptual and semantic properties of the timbre. This challenge is often called the synthesizer programming problem[1], which can even hinder the creative process for experienced musicians and professionals[2].

To address this problem, the concept of extracting synthesizer features has been proposed. [3]–[6]. This approach involves extracting features from a user-supplied sound (the target sound) and providing macro-controls that enable users to adjust the timbre based on these features, thereby simplifying the process of achieving the desired timbre.

Our research employs wavetable synthesis (WTS), one of the earliest software synthesis methods developed by Max Mathews in the late 1950s[7]. WTS, the core technology behind many synthesis methods, remains popular in numerous software synthesizers today. WTS works by cyclically referencing and replicating a waveform (WT: Wavetable) to produce a sound. The timbre is typically modified using filters, amplifiers, or morphing between different wavetables(WTs).

Furthermore, recent years have seen significant advancements in deep generative models. Among these, autoencoder (AE) [8] and variational autoencoder (VAE) [9] have gained prominence in various fields. These models compress the input data into a compact, low-dimensional latent space and reconstruct the original data from this compressed representation, facilitating the extraction of valuable feature representations from the data. Additionally, conditional variational autoencoder(CVAE)[10], which enables conditional generation based on specified conditions, has also been introduced.

Our research uses a CVAE to generate a WT for a single cycle, which is then applied to the oscillator within WTS. We have engineered a model that allows users to define the desired timbre through semantic labels. Figure 1 depicts a schematic overview of this process. In previous studies, methods for generating WT using deep generative models have been proposed[3], [11], [12]. However, issues regarding the diversity, controllability, and accuracy of the generated WT remain to be solved. Diverging from these methods, our research proposes a technique that enables users to emphasize or attenuate specific harmonic overtones within a given WT using semantic labels. This approach represents the primary contribution of our research.

## II. Related Works

### A. Synthesizer Programming Problem

The synthesizer programming problem, as outlined in Section I, is a complex issue that experienced musicians and professionals face. This challenge stems from the intricate control mechanisms and the necessity for specialized expertise, particularly in the realm of synthesizer sound creation [2]. Substantial research has focused on automating the estimation of synthesis parameters to address this issue. As a result, two approaches have emerged: the first aims to approximate synthesis parameters to replicate a target sound [13]–[16] and the second utilizes descriptive language concerning timbre for parameter estimation [17]–[19].

The feature synthesizer methodology warrants particular attention. This method is swift and user-friendly, specifically designed for music production and live performances. It involves extracting features from target timbres and providing macro-controls, thereby facilitating the adjustment of timbre based on these extracted features. This approach significantly simplifies the process of achieving the desired timbre. The efficacy of the feature synthesis method in enabling intuitive timbre modification has been substantiated in various research studies[3]–[5], [11], [20]. Our research focuses on this feature synthesis approach, emphasizing its ability to rapidly adapt to changes in timbre during music production and performance.

---
[1] https://noto.li/XZKK4q

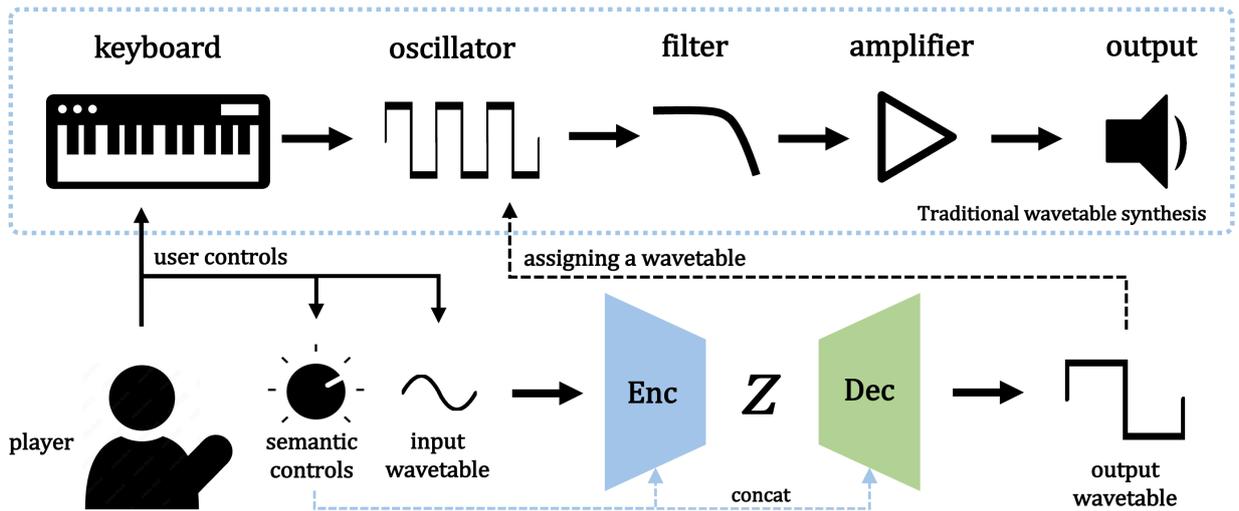

Fig. 1. Schematic showing an overview of the wavetable synthesis (WTS) method demonstrated in this research. A participant creates a wavetable (WT) exhibiting a specific timbre, achieved by conditioning the initial WT with various semantic labels. This generated WT is consistently referenced and emitted as a continuous sound. In this WTS setup, the pitch and duration can be adjusted using a keyboard or similar devices, akin to the conventional WTS operations. Additionally, the timbre of the WT can be modified by using filters and amplifiers.

*B. Neural Audio Synthesis*

Neural audio synthesis is a technology used for generating or transforming audio data using deep learning. As outlined in Section I, deep generative models are rapidly developing and have become an essential technology in music. Research to generate audio data is also active, and research for applications to synthesizers has been proposed[21]–[23].

Several approaches have been proposed to complement the existing synthesis methods using deep learning. To solve the synthesizer programming problem, Esling et al. [24] showed that VAE can simultaneously process parameter estimation, timbre control, and reconstruction with semantic labels in a single model for subtractive synthesis. Kreković [3] illustrated the feasibility of processing and controlling parameter estimation, timbre control, and reconstruction with semantic labels in a unified WTS model, highlighting the generation of WT from semantic labels.

This research explores the application of deep learning to musical synthesis, akin to the earlier research. It focuses on the latest technical trends in model construction, loss functions, and methods for calculating semantic labels.

*C. Advances in Wavetable Synthesis*

Wavetable synthesis generates a sound by repeatedly referencing a waveform for one cycle. This method, widely used in various genres, faces the common challenge of timbre generation, as with other synthesis methods. Creating a WT generally involves editing the waveform and frequency spectrum [25] or choosing a preset WT.

However, while these methods significantly impact timbre, they often diverge from human intuition. For instance, subtle changes in the shape or frequency spectrum of a WT can alter its timbre. Nonetheless, discerning the characteristics of the timbre through visual inspection is challenging. Even when selecting a WT from a preset, users must remember the relationship between the name of the preset and its corresponding timbre, which requires a certain level of expertise. Consequently, achieving the desired timbre in WTS can be difficult.

To address this problem, as outlined in Section I, previous research generated WT using VAE[12] and AE[11], and the timbre was changed by randomly manipulating the latent space. Furthermore, Kreković [3] proposed a model that generates WT from three input labels: $bright$, $warm$, and $rich$, using a decoder [26]. However, embedding WT features into a limited number of latent spaces (labels) poses challenges. The diversity and accuracy of the generated WT are still two issues that remain to be solved. To enhance the diversity and accuracy of the WT that can be generated, we have proposed a model that uses CVAE[10]. This model is designed to create WT that exhibits a one-period waveform characteristic.

### III. METHODS

*A. CVAE*

The CVAE[10], an extension of the VAE, is a generative model. This model is notably effective in generating data conditioned on a specific variable, denoted as $c$. In our research, we incorporated the condition, $c$, into the encoder and the decoder to apply these conditions effectively.

The input waveform, $x$, and condition, $c$, (in this research, WT and semantic labels, respectively) were mapped to the latent variable, $z$, by the encoder, $q_\phi(z|x,c)$. Subsequently, the decoder, $p_\theta(x|z,c)$, generated the output data, $\hat{x}$.

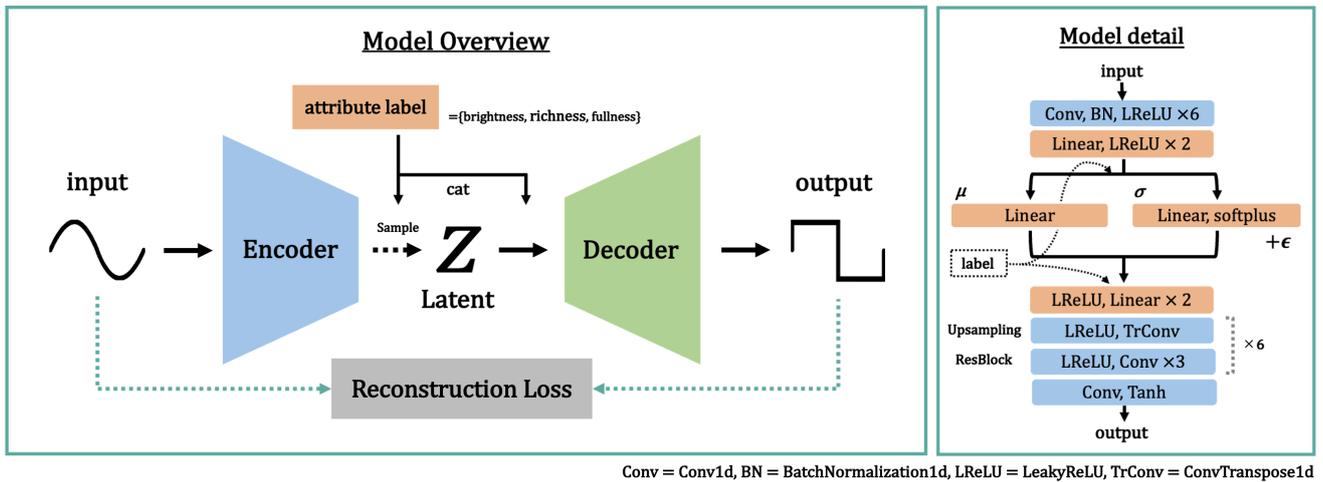

Fig. 2. Schematic of the model configuration. Left: Model overview, right: detailed configuration.

## B. Loss Function

The loss function comprised the reconstruction loss and the Kullback–Leibler (KL) divergence term. These components were optimized by minimizing them through the stochastic gradient descent (SGD) method. The loss function in this research was defined as follows:

$$L_{\text{CVAE}}(x) = \mathbb{E}_{\hat{x} \sim p_\theta(x|z)}[d(x, \hat{x})] + \beta \times D_{\text{KL}}[q_\phi(z|x,c) || p(z)] \quad (1)$$

The first term in Equation (1), the reconstruction loss, was used for evaluating the accuracy of the generated data in reconstructing the original data and aimed to minimize the information loss between the encoder and decoder. Previous research has suggested using multiscale spectral distance to estimate the waveform distances[21], [23]. Due to the short data length in our WT approach, we calculated the spectral distance using only the maximum table size (3600 samples in our research) after iterative concatenation.

The function, $d$, for the reconstruction loss was defined as follows:

$$d(x, \hat{x}) = \frac{||S_x - S_{\hat{x}}||_F}{||S_x||_F} + \log ||S_x - S_{\hat{x}}||_1 \quad (2)$$

where $S_x$ and $S_{\hat{x}}$ are the amplitude spectra of $x$ and $\hat{x}$, respectively. Here, $||\cdot||_F$ denotes the Frobenius norm and $||\cdot||_1$ the L1 norm.

The amplitude spectrum of the input waveform, $x$, and the output waveform, $\hat{x}$, were compared and their loss was calculated. This approach is based on the assumption that the perceptual characteristics of the synthesized sound remain almost invariant between the signal waveforms that are out of phase. Following the convention adopted by Kreković [3], we improved the frequency resolution of the spectrum by concatenating six WTs iteratively, which achieved the best performance in the investigation stage in terms of reconstruction error and controllability of conditioning.

The second term of Equation (1), the KL divergence term, measured the closeness of the latent variable, $z$, sampled from the encoder, $q_\phi(z|x,c)$, to the predefined multivariate normal distribution, $p(z)$. This term regularized the latent space, ensuring that the model captured the general data characteristics instead of merely memorizing the data.

The hyperparameter, $\beta$, adjusted the balance between the reconstruction loss and the KL divergence term. The setting of this parameter significantly influenced the performance of the model, as proposed by Higgins et al. [27].

## C. Model Configuration

The model's architecture was adapted from a structure outlined in previous research[23] and was modified to enable the conditional generation of WT. The encoder combined the convolutional layers that convert the WT into a 32-dimensional latent representation. The decoder used an upsampling layer and a residual network[28] to generate the WT. Conditional generation was executed by concatenating the relevant labels to the inputs and outputs within the latent space[12]. Figure 2 illustrates an overview of the model structure.

## D. Computation of Semantic Labels

Previous research has explored the computation of semantic labels in sounds, including WTS. In our research, we employed the labels $bright$, $warm$, and $rich$, following the methodology outlined by Kreković [3], [29]. These labels were derived from the spectral centroid, the energy ratio of the odd harmonics, and the spectral density, followed by normalization.

The spectral centroid, representing the center of mass of the spectrum, indicates the balance between the high and low frequencies and correlates with the timbre brightness[30]. The energy ratio of the odd harmonics, i.e., the proportion of energy in the odd harmonics compared to other harmonics, was associated with the warmth of the timbre[31]. The spectral density, calculated as the spectrum's standard deviation, corresponded to the richness of the timbre[32].

Six WTs were concatenated in succession to improve spectral resolution, following the method outlined in Section III-B

for extracting features from a single WT. Audio features were computed using a fast Fourier transform and a Hanning window. Since the derived values do not naturally correspond with human perception, a transformation was applied to ensure linear alignment with human perception, particularly within the range of [0, 1]. However, the energy ratio of the odd harmonics for the $warm$ label, inherently falling within [0, 1], required no additional normalization.

## IV. EXPERIMENTS

### A. Dataset

We utilized 4158 WT data from Adventure Kid Research & Technology[2] for the dataset. For the size of the sample for each WT, we select 600, which is larger than that utilized in previous works (i.e., 320 in [11], 327 in [3], and 512 in [12]. ), to enhance the representativeness of high-frequency components in WTS. The dataset was partitioned into three segments: 80 % for training, 10 % for validation, and 10 % for testing. This distribution ensured a comprehensive evaluation of the model across different data subsets.

### B. Training

Training was conducted by feeding the model with training data and the corresponding labels. The Adam optimizer[33] was used with a learning rate 1e-4. The batch size was 32, and the number of epochs was 30000 to minimize the reconstruction and KL divergence.

Over-optimizing the KL divergence between the prior and posterior distributions in the early stages of training can lead to an issue known as KL vanishing[34], in which the latent variables are ignored in the reconstruction. To mitigate this issue, the variable, $\beta$, for the KL divergence was gradually increased according to the following Equation:

$$\beta(e, w, \beta_{\min}, \beta_{\max})$$
$$= \begin{cases} \beta_{\max} & \text{if } e > w \\ \exp\left(t \cdot (\log \beta_{\max} - \log \beta_{\min}) + \log \beta_{\min}\right) & \text{otherwise} \end{cases} \quad (3)$$
$$t = \frac{e}{w} \quad (4)$$

where $e$ is the current number of epochs, $w$ is the upper bound on the number of epochs, $\beta_{\min}$ is the minimum value set for $\beta$, $\beta_{\max}$ is the maximum value set for $\beta$, starting from $\beta_{\min} > 0$ and increasing exponentially to $\beta_{\max}$ exponentially as the learning progresses.

We empirically set $w = 10000$, $\beta_{\min} = 1e - 4$, $\beta_{\max} = 1e - 1$, and 1 epoch for the period of the increasing.

## V. EVALUATION

The reconstruction quality of the generated WT and the controllability and generation time of the conditioning generation were evaluated. This research aimed to enhance the intuitiveness of timbre control and enable real-time timbre searches. To achieve this, the following three steps were essential: 1) high-quality reconstruction of the WT input to the model 2) conditional generation according to the values of the semantic labels, and 3) short processing time for generation and real-time use. The results obtained by performing the abovementioned steps are shown below. The source code of the model and the weights of the trained model used to obtain the results are also available[3]

### A. Reconstruction Quality

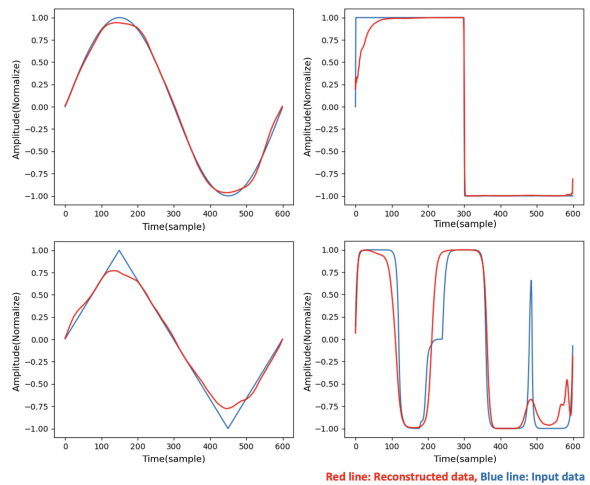

Fig. 3. Example of the qualitative evaluation of the reconstruction quality. The plots show that the reconstruction is done according to the input data.

The mean absolute error (MAE) was calculated to evaluate the reconstruction quality between the reconstructed waveform and the original input waveform. The calculated MAE amounted to 0.082 for the waveform state and 0.013 for the logarithmic amplitude spectrum. The logarithmic amplitude spectrum was calculated by repeatedly concatenating the six WTs as described in Section III-B to facilitate a comparison between the frequency components. The reconstructed waveforms are shown in Figure 3, which reproduce the input waveforms.

These results indicate that the reconstructed waveform successfully emulates the input waveform, capturing the essential characteristics of the WT. Future work will entail a comparison of our method with other models and conducting listening experiments to validate the reconstruction quality quantitatively.

### B. Controllability of conditioning generation

To evaluate the controllability of the conditioning generation, the MAE and Pearson correlation coefficients of the output labels concerning the input labels (condition $c$) were checked. The output labels were computed from the output waveforms using the same procedure as that described in Section III-D. The higher number of minor differences between the input and output labels and higher correlation coefficients indicate better control of conditioning generation. Figure 4

---

[2]https://www.adventurekid.se/akrt/

[3]https://github.com/tsugumasa320/WavetableCVAE

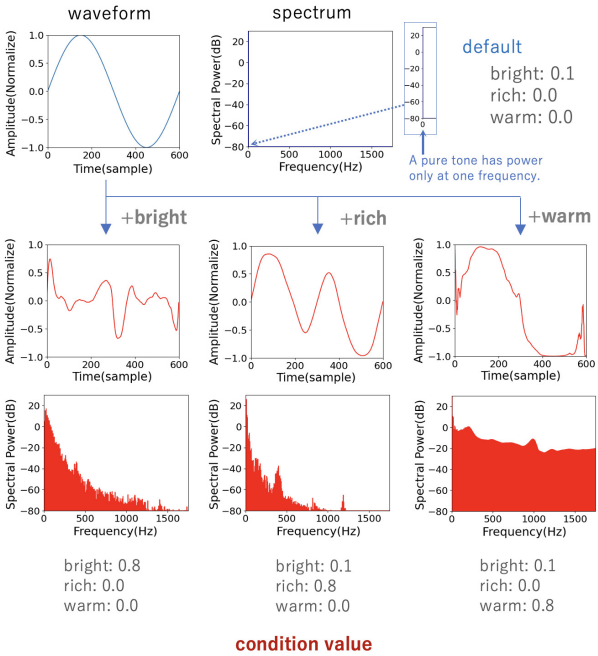

Fig. 4. Outcomes of conditional generation using sine waves as inputs: an analysis based on the Label characteristics. $bright$ refers to the spectral centroid, $rich$ denotes the energy ratio of the odd harmonics, and $warm$ corresponds to the spectral density. The value associated with each label increases with the value of the conditioning.

illustrates a section of the results from the conditioning generation process.

The calculation procedure was as follows: A single input waveform, $x$, and an input label, $c$, from the dataset were selected. The input label, $c$, was a variable that comprised three precomputed values: $bright$, $warm$, and $rich$. From $c$, one evaluation label, $c_e$, was selected to evaluate the controllability of conditioning generation. The label $c_e$ was set in order of 20 equally spaced values in the range of [0, 1]. Labels that were not used for evaluation were also needed for conditioning generation. The labels obtained from the input waveforms were calculated using the same procedure as in Section III-D.

The output waveform, $\hat{x}$, and the evaluation label, $\hat{c}_e$, were obtained from the input waveform, $x$, and the input label, $c$. This procedure was repeated for the number of divisions of $c_e$ used as input (20 times in this research).

The transition between $c_e$ and $\hat{c}_e$ was stored as an array of 20 points. The MAE and Pearson correlation coefficients were calculated and evaluated from the stored arrays. These processes were repeated for WT, and the averages were obtained with the three parameters, $bright$, $warm$, and $rich$, for the entire dataset.

The value of MAE averaged over $bright$, $warm$, and $rich$ were 0.19, 0.05, and 0.14, respectively. The $warm$ value means the change is consistent with the label, whereas $bright$ and $rich$ remain challenging. The distribution of the training labels may have been biased, limiting the exposure of data to a sufficient range of labels for practical training.

The averaged values of the Pearson correlation coefficients were 0.97 for $bright$, 0.98 for $warm$, and 0.96 for $rich$. Strong positive correlations were confirmed for all parameters. This indicates that conditioning generation responds appropriately to the increase or decrease of each parameter.

*C. Generation Time*

The time required for generating WT was evaluated to ascertain the real-time performance capabilities of the proposed method. WT was generated 100 times, and the average generation time was calculated to be approximately 2.7 ms. The evaluations were conducted on a MacBook Pro (13-inch, M1, 2020) with an M1 Chip (CPU), utilizing PyTorch Lightning[4] 1.7.7. This notably brief generation time suggests the feasibility of real-time operation of our method using only the CPU without requiring a specialized GPU for deep learning.

## VI. Summary and Future Issues

In this research, we proposed a method for semantic timbre control of WT using CVAE, and demonstrated the possibility of improving the synthesizer programming problem. The following is the rationale for the proposed method:

This method demonstrates that the reconstructed waveform can accurately follow the input waveform by effectively capturing the timbre features of the WT. Additionally, we have confirmed that conditioning generation can be controlled according to the intended outcome[5]. However, the inherent complexity of timbre variation indicates that more refined control is essential. Future research should explore increasing the number of conditioning labels to facilitate more nuanced sound manipulation, thereby contributing to the resolution of the synthesizer programming problem.

Research examining the application of deep learning to WTS is difficult to evaluate in a unified manner due to the variety of model architectures and research objectives[3], [11], [12]. As outlined in Section II-C, future research will need to develop quantitative metrics to assess the quality of waveform reconstruction and the controllability of conditioning, enabling better comparisons across different models and providing a more coherent basis for advancing WTS research..

Furthermore, as detailed in Section V-B, the controllability assessment of the conditioning generation indicates significant MAE, particularly with $bright$ and $rich$ labels. The bias in training label distribution is likely the contributing factor, necessitating countermeasures. Results may be enhanced by augmenting the dataset size, altering label analysis, and normalization methods. Another strategy could involve adopting latent space and label feature separation mechanisms akin to those in Fader Networks [35].

With regards to the generation time, the results reveal that real-time use is possible using only the CPU without needing a particular GPU for deep learning.

As for real-time performance, model weight reduction methods such as model parameter quantization and branch pruning could further reduce the processing time of the proposed method.

---

[4] https://lightning.ai/docs/pytorch
[5] The reconstruction accuracy is shown in Figure 3, and the controllability of conditioning generation is shown in Figure 4